\documentclass[times]{mn2e}
\usepackage{times}
\input epsf.sty

\def\rQPO{r_{\rm QPO}}

\def\ti{t_{\rm i}}
\def\Ka{K$\alpha$\ }
\def\gmin{g_{\rm min}}
\def\gmax{g_{\rm max}}
\def\kT{k T_{\rm e}}
\def\Ts{T_{\rm s}}
\def\tauT{\tau_{\rm T}}

\title[Energy spectra of high-frequency QPO]
{Modelling the energy dependencies of high-frequency QPO in black hole X-ray 
binaries}

\author[P. T. \.{Z}ycki, A. Nied\'{z}wiecki and M. A. Sobolewska]
{Piotr T. \.{Z}ycki$^1$\thanks{e-mail: ptz@camk.edu.pl}, Andrzej Nied\'{z}wiecki$^2$ 
    and Ma{\l}gorzata A. Sobolewska$^3$\\
    $^1$Nicolaus Copernicus Astronomical Center, Bartycka 18, 00-716 Warsaw, Poland \\
    $^2${\L}\'{o}d\'{z} University, Department of Physics, Pomorska 149/153, 90-236 {\L}\'{o}d\'{z}, Poland \\
    $^3$University of Durham, Physics Depertment, South Road DH1 3LE, UK}

\date{Accepted 2007 April 24. Received 2007 April 13; in original form 2007
January 09}

\begin{document}
\label{firstpage}

\maketitle

\begin{abstract}

We model energy dependencies of the quasi periodic oscillations (QPO) in the
model of disc epicyclic motions, with X-ray modulation caused by varying
relativistic effects. The model was proposed to explain the high frequency
QPO observed in X-ray binaries. We consider two specific scenarios for the
geometry of accretion flow and spectral formation. Firstly, a standard cold 
accretion disc with an active X-ray emitting corona is assumed to oscillate. 
Secondly, only a
hot X-ray emitting accretion flow oscillates, while the cold disc is absent
at the QPO radius. We find that the QPO spectra are generally similar to the
spectrum of radiation emitted at the QPO radius, and they are broadened by the
relativistic effects. In particular, the QPO spectrum contains the disc
component in the oscillating disc with a corona scenario.
We also review the available data on energy dependencies of high frequency QPO,
and we point out that they appear to lack the disc component in their
energy spectra. This would suggest the hot flow geometry in the spectral states
when high frequency QPO are observed.

\end{abstract}


\section{Introduction}

High frequency (hecto-Hz) quasi-periodic oscillations are observed in 
X--ray power spectra of  a number of black hole X-ray binaries. 
In some sources they occur in pairs, with frequency ratios close to 3:2
(see Remillard \& McClintock 2006 for a recent review and Lachowicz, Czerny
\& Abramowicz 2006 for similar analysis of active galactic nuclei data).
This latter property of the high-$f$ QPO (hfQPO hereafter) has 
stimulated  development of a class of models, which interpret
the QPO as a result of non-linear resonances (Klu\'{z}niak \& Abramowicz 2001).
Since the QPO frequencies are close to dynamical frequencies very near the
central black hole, the models involve fast dynamical processes, for example
epicyclic motions. In particular, resonances between the radial and vertical
epicyclic motions have been considered by Abramowicz et al.\ (2003).
Another suggestion was made by Blaes, Arras \& Fragile (2006), who proposed 
the ``breathing mode'' (vertical pulsations) of a radiation pressure accretion 
disc in resonance with  vertical epicyclic motion.
The resonance condition $\nu_\theta/\nu_r=3/2$ then selects the radius where 
the oscillations are concentrated, $\rQPO$, 
and the observed modulation is produced. This radius
has to be smaller than $6 GM/c^2$ to match the values of frequencies 
actually observed in black hole binaries, indicating rotating
black holes (e.g., T\"{o}r\"{o}k et al.\ 2004).

The actual modulation of X-rays in the model of radial/vertical epicyclic 
motion resonances could be produced by  variable relativistic
effects, as the oscillating disk changes its position relative to un-perturbed
position (Bursa et al.\ 2004). These effects obviously affect only spectral
components emitted at the QPO radius.

Generally, X-ray emission from accreting black holes consists 
of two main spectral components: thermal emission from the accretion disc and 
a component produced by inverse Compton up-scattering of a fraction
of the disc photons in a hot plasma.  In some sources a more complex
spectra are also observed, e.g., with the disc component and two Comptonized
components (e.g., Gierli\'{n}ski \& Done 2003 and references therein).  
The reprocessed component (iron fluorescent
K$\alpha$ line and the Compton reflected continuum) are often present as well.
The quasi-periodic X-ray modulation may then appear in all or only some of 
the components. Determining which components are actually modulated
provides an important constraint on the model of the oscillations, since
it constrains the physical process actually modulating the X-rays.

For example, energy dependencies of a related QPO phenomenon, the low-frequency
QPO ($f\la 10$ Hz) in black hole binaries, reveal that it is the Comptonized
component that is modulated (Sobolewska \& \.{Z}ycki 2006).
The disc component does not participate in the modulation. Moreover,
the QPO spectrum is harder than the time averaged spectrum (at least
in the soft spectral states), consistent with the QPO being driven by
modulations of the heating rate of the comptonizing plasma.
This has an important consequence in that any model postulating disc
origin of the oscillations must provide a mechanism of transferring the
oscillations to the hot Comptonizing plasma without affecting the
disc emission (\.{Z}ycki \& Sobolewska 2005).

In this paper we attempt to make a first step towards similar analysis
of the hfQPO. First, we review the available data on energy dependencies
of the hfQPO. We then simulate the dependencies for the 
specific hfQPO model of resonant epicyclic motions, with X-ray modulations
solely due to varying relativistic effects.

We consider two specific geometrical models.  In the first model the standard
accretion disc is envisioned to extend all the way to the last stable orbit and
the hard X-rays are produced in a hot corona extending above the disc. 
As a consequence, three spectral components are produced at the QPO radius:
thermal disc emission, hard Comptonized component and the reprocessed
component. We will also extend this model to include an additional 
optically thick Comptonized component emitted at the QPO radius. This is 
motivated by results of more careful analyses of spectra and energetics of 
very high state showing such additional component (e.g., Done \& Kubota 2006
and references therein). In this scenario we assume that both the disc and the corona
undergo epicyclic oscillations, so the relativistic effects affect all spectral
components.

In the second model, the standard disc is truncated somewhere above $\rQPO$ and 
replaced by an inner hot flow. Thus, only the hot plasma undergoes epicyclic
oscillations, while the disc flux is not subject to those motions. 
Only the Comptonized component is thus produced at 
$\rQPO$, and it is modulated by the relativistic effects.

\section{Observational data of high frequency QPO}
\label{sec:data}

\begin{figure}
 \parbox{0.5\textwidth}{
  \epsfxsize = 0.45\textwidth
  \epsfbox[18 150 620 710]{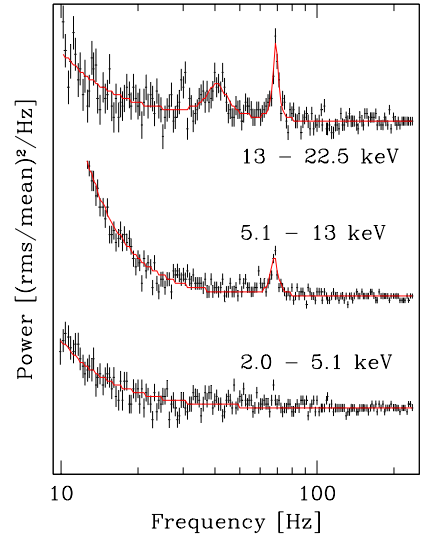}
}
 \caption{High-$f$ part of power spectrum of GRS~1915+105 in three energy bands 
(white noise was not
subtracted and the spectra were shifted with respect to each other for a
better comparison). In the softest energy band the high-$f$ QPO at 67 Hz is
undetectable. It appears in the intermediate energy band with rms of
0.9 per cent. In the hardest energy band the rms of the QPO increases to
3.5 per cent. In addition, another 42 Hz QPO appears. The frequencies of these two
QPO are in a ratio close to 3:2.
\label{fig:grshfqpo}}
\end{figure}

To date there are seven X-ray black hole binaries with power spectra
that show high-frequency ($ \ga 100$ Hz) QPO.
Usually, these fast QPO are detected in the hard energy
band and are less significant in the soft energy band
(e.g.\ 4U~1630-47: Klein-Wolt, Homan \& van der Klis 2004;
GRS~1915+105: Belloni et al.\ 2006). 
For instance, in XTE~J1859+226 oscillations with frequencies of 150 and 
187 Hz were found in the 5.9--60 keV band, while in the 2--5.9 keV band 
neither of the QPO yielded a significant detection  (Cui et al.\ 2000). 
In XTE~J1650-500 the 250 Hz QPO is detected with rms amplitude of 
$<0.85$ per cent, 4.5 per cent, and $<12.1$ per cent in the 2--6.2, 
6.2--15.0 and 15.0--60.0 keV bands, respectively (Homan et al.\ 2003).
In XTE~J1550-564 the 185 Hz QPO is found with rms amplitude significantly 
higher in the 6--12 keV band than in the 2--6 keV band 
(Remillard et al. 1999a; see also Miller et al.\ 2001 and Homan et al.\ 2001).
In GRO~J1655-40 the 450 Hz QPO shows an increase of the rms amplitude 
from $<0.4$ per cent in the 2--12 keV band to 4.8 per cent in the
13--24 keV band (Strohmayer 2001a).
In  H~1743-322 the 160 and 240 Hz QPO have rms increasing from $<0.6$ per cent 
($<$1.2 per cent) in the 3.3--5.8 keV band to 2.3 per cent
(1.3 per cent) in the 5.8--20.9 keV band for upper (lower) QPO and both QPO 
have rms amplitude $<$13 per cent in the 21--60 keV band
(Homan et al.\ 2005; see also Remillard et al.\ 2006).
The 67 Hz QPO in GRS~1915+105 has the rms amplitude increasing from 1.5 per 
cent below 5 keV to 6 per cent above 13 keV (Morgan, Remillard \& Greiner 1997; 
Cui 1999). To illustrate the trends in energy we plot in Fig.~\ref{fig:grshfqpo} 
the high-$f$ power spectra of GRS~1915+105 data analyzed in Strohmayer (2001b).

There are however observations that contradict this general trend.
For example, the slightly slower 65 Hz QPO discovered in XTE~J1550-564 shows 
very clear decrease of the rms amplitude with energy
(5.5 per cent in the 2--6 keV band, 4.7 per cent in the 6--15 keV band, and
$<4.5$ per cent in the 15--67 keV band; Kalemci et al.\ 2001). We note
though that this QPO was discovered  during a transition to the hard state, 
while the other hfQPO are typically seen in the very high state. 
Another interesting example is that of GRO~J1655-40. While the 450 Hz QPO shows 
the typical increase in the rms amplitude as mentioned above, 
the 300 Hz QPO is seen {\em only\/} in the soft, 2--12 keV, band with typical rms 
amplitude of 0.8 per cent (Remillard et al.\ 1999b; Strohmayer 2001a). 
This could mean that the 300 Hz QPO has a different energy dependence and its 
rms decreases with energy. Another possibility is that the rms is simply too low 
to be 
detectable in the hard band, where photon statistics is worse.
This was pointed out by Strohmayer (2001a) who place the upper limit on 
the QPO rms amplitude in the 13--24 keV band at 1.7 per cent.  

The  interesting phenomenon of double hfQPO with integral frequency ratios of 
2:3 or 3:5 was observed in a few sources. These include the 300 and 450 Hz QPO
in GRO~J1655-40 (Remillard et al.\ 1999b, 2002; Strohmayer 2001a), 
160 and 240 Hz QPO in H~1743-322 (Homan et al.\ 2005; Remillard et al.\ 2006),
150 and 187 Hz QPO in XTE~J1859+226 (Cui et al.\ 2000), or somewhat
lower frequency, 41 and 67 Hz, QPO in GRS~1915+105 (Morgan et al.\ 1997;
Strohmayer 2001b). However, these pairs of QPO are not always  detected
simultaneously. Moreover, due to limited statistics it is not always
apparent whether this is a pair of QPO, or one QPO moving in frequency
(e.g.\ Miller et al.\ 2001; Homan et al.\ 2003). The best example here
may be XTE~J1550-564 for which the hfQPO were observed in a wide
frequency range, 100--285 Hz (e.g.\ Homan et al.\ 2001). However,
Remillard et al.\ (2002) show that frequencies of these QPO cluster 
preferentially around 184 and 276 Hz.

\section{The Model}
\label{sec:model}

We adopt the standard description of overall X-ray emission from X-ray 
binaries,
that the thermal emission comes from an optically thick accretion disc, and 
the disc photons can be Comptonized in a hot plasma forming the hard emission
component. The hard X-rays can be further reprocessed in the accretion disc.
Our basic assumption is that there is no specific quasi-periodic modulation of 
the intrinsic emission, and the observed QPO is a result of
relativistic effects (although the intrinsic emission can be randomly
variable producing the usual broad band power spectrum).
We consider two geometrical scenarios: an accretion disc extending to the last 
stable orbit with a hot active corona above it, and an accretion disc replaced
by a hot flow inside certain radius, somewhat larger than the QPO radius. 

Emission from the optically thick accretion disc is described by the standard
disc black body approximation with temperature varying with radius as,
\begin{equation}
 T(r) = T_0 F(r)^{1/4},
\end{equation}
where $F(r)$ (assumed normalized to unity at its maximum) describes the radiation
flux due to gravitational energy dissipation, as a function of radius for a 
given angular momentum of the black hole (see Krolik 1999, p.\ 150). 
The Comptonized component is computed using the {\sc thComp} code of
Zdziarski, Johnson \& Magdziarz (1996). This is parameterized by the asymptotic 
slope of the spectrum, $\Gamma$, plasma optical depth, $\tauT$, and seed photon
temperature $\Ts$. We assume that hot plasma properties do not depend on the 
radius, and we adopt $\Gamma=2.25$, $\tauT=0.1$. Obviously, $\Ts = T(r)$.
The reprocessed component is represented here by the Fe K$\alpha$ line emitted
at 6.4 keV, with equivalent width of 100 eV.

For photon transfer we consider a Kerr black hole characterized by its mass, 
$M$, and
angular momentum, $J$. We use the Boyer-Lindquist coordinate system 
$(t,R,\theta,\phi)$. The following dimensionless parameters are used below
\begin{equation}
r = {R \over R_{\rm g}},~~~\hat t = {ct \over R_{\rm g}}, 
~~~a = {J \over c R_{\rm g} M},
~~~\Omega = {d\phi \over {d{\hat t}}},
\end{equation}
\begin{eqnarray}
 \Delta & = &r^2-2r+a^2 \\
\Sigma & = & r^2 + a^2 \cos^2 \theta \\
 A & = & (r^2+a^2)^2 - a^2 \Delta \sin^2 \theta, 
\end{eqnarray}
where  $R_{\rm g} = GM/c^2$ is the gravitational radius. 

We assume that a disc element oscillates vertically and radially with respect 
to an equilibrium circular orbit
\begin{eqnarray}
r(\hat t) & = & r_0 + r_{\rm max} \cos (\Omega_r \hat t) \label{equ:oscradial}\\
\theta(\hat t) & = & \theta_0 + {r_{\rm max} \over r_0} 
\cos (\Omega_{\theta} \hat t)
\label{equ:oscillation}
\end{eqnarray}
where $\Omega_{\theta}$ and $\Omega_{r}$ are vertical and radial epicyclic 
frequencies observed at infinity:
\begin{eqnarray}
\Omega_{\theta}^2 & = & \Omega_{\rm K}^2 
  \left( 1 - 4 a r^{-3/2} + 3 a^2 r^{-2} \right) \\
\Omega_r^2 & = & \Omega_{\rm K}^2 
\left( 1 - 6 r^{-1} + 8 a r^{-3/2} - 3 a^2 r^{-2} \right), 
\end{eqnarray}
$\Omega_{\rm K}$ is the Keplerian angular velocity
\begin{equation}
\Omega_{\rm K}= {1 \over r^{3/2}+a}.
\end{equation}
The equilibrium position corresponds to $\theta_0=\pi/2$ and $r_0$
given by the resonance condition $\Omega_{\theta}/\Omega_r = 3/2$.
This gives $r_0=3.9$ for $a=0.998$ and $r_0=10.8$ for $a=0$.
In our simulations we assume $r_{\rm max} = 0.1 r_0$.
Note that we assume that $r({\hat t})$ and $\theta({\hat t})$ do not 
depend on the 
azimuthal angle, which means that the whole ring oscillates in phase.

We construct the transfer function following a Monte Carlo method described 
in detail in \.Zycki \& Nied\'zwiecki (2005), modified to implement 
the source motion corresponding to oscillations.  A large number
of photons are emitted from consecutive positions of the source, 
$r_i = r(\hat t_i)$ and $\theta_i= \theta(\hat t_i)$ given by equations 
(\ref{equ:oscradial}) and
(\ref{equ:oscillation}), with the phase  of oscillations parametrized 
by $\hat t_i$. The initial azimuthal angle in the source rest frame, 
$\phi_{\rm em}$, is generated from the uniform distribution. 
The polar angle between the  photon initial direction and the vertical direction,
$\theta_{\rm em}$, is used as a parameter for the transfer function. 
Solving equations of photon motion, as described below,  we find the energy 
shift ($g = E_{\rm inf}/E_{\rm rest}$; $E_{\rm inf}$ and $E_{\rm rest}$ is the 
photon energy at infinity and in the source rest frame, respectively), 
inclination and arrival time, 
($g$, $i$, $\Delta \hat t$). Each element of  the transfer function, 
${\cal T}(\hat t_i, i, g,\Delta \hat t, \theta_{\rm em})$, 
is computed by summing all photon trajectories  emitted from $(\theta_i,r_i)$ 
for all angles, $\phi_{\rm em}$, for which required ($g$, $i$, $\Delta \hat t$) 
are obtained.

The travel time and the change of azimuthal angle of a photon emitted at $(\theta_i,r_i)$
are given by 
\begin{equation}
\Delta \phi_{\rm ph} =  \int\limits_{\theta_i}^i {\lambda - a \sin^2 \theta \over 
  \Theta^{1/2} \sin^2 \theta  } {\rm d} \theta + \int\limits_{r_i}^D
{a(r^2 + a^2 - \lambda a) \over \Delta \Re^{1/2} } {\rm d}r,
\end{equation}
\begin{equation}
\Delta \hat t_{\rm ph} =  \int\limits_{\theta_i}^i {a(\lambda - a \sin^2 \theta) \over 
  \Theta^{1/2} } {\rm d} \theta  + \int\limits_{r_i}^D
{(r^2 + a^2)(r^2 + a^2 - \lambda a) \over \Delta \Re^{1/2} }  {\rm d}r,
\end{equation}
where $\Re$ and $\Theta$ are radial and polar effective potentials, respectively,
\begin{eqnarray}
\lefteqn{\Re(r) = (r^2+a^2-\lambda a)^2 - \Delta \left[ \eta + (\lambda - a)^2
\right] ,} \nonumber \\
\lefteqn{\Theta(\theta) =\eta + \cos^2 \theta \left( a^2-\lambda^2/{\sin^2 
\theta} \right),}
\end{eqnarray}
$\eta$ and $\lambda$ are photon  constants of motion, 
\begin{equation}
\eta \equiv {Q c^2 \over E_{\rm inf}^2 R_{\rm g}^2},~~~
\lambda \equiv {L c \over E_{\rm inf} R_{\rm g}},
\end{equation}
$Q$ is the
Carter's constant and $L$ is the component of angular momentum parallel
to the black hole rotation axis.
The angle, $i$, at which the photon is observed far from the source  
is determined by the integral equation of motion
\begin{equation}
\int_{r_i}^D \Re^{-1/2}{\rm d} r=\int_{\theta_i}^{i} \Theta^{-1/2} {\rm d}\theta.
\label{int}
\end{equation}

Photon momentum components in the source rest frame are related to the emission angles,
$p_{\theta} = E_{\rm rest} \cos \theta_{\rm em}$, $p_{\phi} = E_{\rm rest} \sin \theta_{\rm em} \sin \phi_{\rm em}$, $p_r = E_{\rm rest} \sin \theta_{\rm em} \cos \phi_{\rm em}$.
Performing two subsequent Lorentz transformations, first from the source rest frame 
to the rest frame of a Keplerian observer, with the relative velocity 
\begin{eqnarray}
v^r & = & - r_{\rm max} \Omega_{(r)} \sin (\Omega_r \hat t_i) \\
v^{\theta} & = & - r_{\rm max} \Omega_{(\theta)} \sin (\Omega_{\theta} \hat t_i) \\
v^{\phi} & = & 0,
\end{eqnarray}
where
\begin{equation}
\Omega_{(\theta)} = \Omega_{\theta} \left( d\tau / d t \right)^{-1}~~~~~~~~~
 \Omega_{(r)} = \Omega_{r} \left( d\tau / d t \right)^{-1}
\end{equation}
and $d\tau / d t$ is the time dilation factor
\begin{equation}
 {d\tau \over {d t}} = r \left( {\Delta \over A} \right)^{1/2} (1-V^2)^{1/2},
\label{equ:dtdt}
\end{equation}
and the second transformation to the locally non-rotating frame (see Bardeen, Press \& 
Teukolsky 1972), with the relative (Keplerian)
velocity
\begin{equation}
 V =  (\Omega_K - 2ar/A) {A \over {r^2 \Delta^{1/2}}}, 
\end{equation}
we find photon energy $E_{\rm ln}$ and momentum $(p_{(r)},p_{(\theta)},p_{(\phi)})$
in the locally non-rotating frame.
Then, photon energy at infinity and constants of motion are given by [see eqs.~(10)
in Nied\'zwiecki (2005)]
\begin{eqnarray}
E_{\rm inf} & = & \left( {\Delta \Sigma \over A} \right)^{1/2} E_{\rm ln} 
+ {2 a r \sin \theta \over (A \Sigma)^{1/2} } p_{(\phi}) \\
\lambda & = & A \left( { E_{\rm ln} \Sigma \Delta^{1/2} \over p_{(\phi)} \sin \theta }
+ 2 a r \right)^{-1} \\
\eta & = & \Sigma { p_{(\theta)}^2 \over E_{\rm inf}^2 } + \cos^2 \theta \left(
{ p_{(\phi)}^2 \over E_{\rm inf}^2 } { A \over \Sigma} -a ^2 \right).
\end{eqnarray}

The transfer function was computed for three periods of the $\theta$-oscillation,
$P_\theta = 2\pi/\Omega_\theta$, $T = 3 P_\theta = 2 P_r$, with the
total time $T$ divided into $N=216$ time intervals. Obviously, the
transfer function is periodic, with period $T$.
The oscillations were followed for a number of $T$ intervals, so that the 
QPO frequencies are better resolved in PDS, and we have checked that the 
final QPO spectra do not depend on the number of $T$ intervals used. 
For the spectral computations presented we simulate the oscillations for total time 
$16 T$.
The time step in the simulations is equal to the time interval between the 
consecutive transfer functions,  $T/N$.
The middle position of the ring is computed at each time point when
a transfer function is known, $\ti$ (modulo $T$), and it is given
by $r(\ti)$ and $\theta(\ti)$ according to equations (\ref{equ:oscradial}) and
(\ref{equ:oscillation}).
For each $\ti$ the energy spectrum of radiation emitted at $\rQPO$ is convolved
with the transfer function producing a signal observed at infinity.

The sequence of spectra created by the above procedure is subject to standard 
analysis in the time and Fourier domains (\.{Z}ycki 2003 and references 
therein).
Power spectra are computed in each energy channel. These consist of a number 
of narrow peaks, since the lightcurves are strictly periodic. 
We compute the rms variability (normalized to the mean) 
integrating each peak over frequency, 
in each energy channel. This, multiplied 
by the time average energy spectrum gives the QPO energy spectrum
(\.{Z}ycki \& Sobolewska 2005). Note, that because of arbitrary values of
epicyclic motion amplitude and a schematic geometry, the rms amplitude of
the model QPO does not have a physical meaning. The purpose of this paper
is to study the shape of the QPO energy spectra.

\section{Results}
 \label{sec:results}

\begin{figure}
 \parbox{0.5\textwidth}{
  \epsfxsize = 0.45\textwidth
  \epsfbox[18 150 620 710]{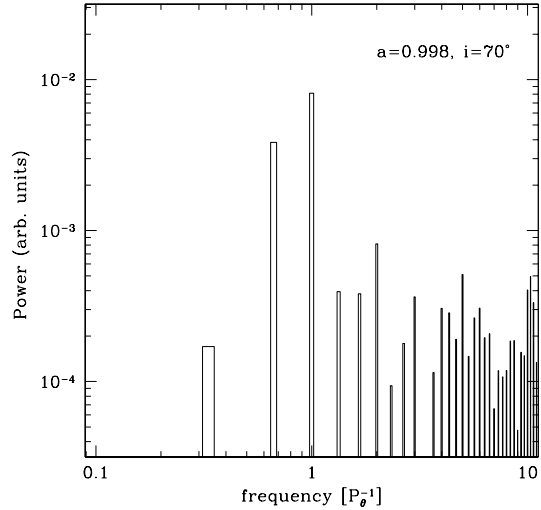}
}
 \caption{Power spectra from the model with maximally rotating, $a=0.998$,
Kerr black hole, for a source inclined at $i=70^\circ$. The two main peaks
correspond to $f=2/3$ and $1 \times P_{\theta}^{-1}$ oscillations. 
Sub(harmonics) are present but are much weaker than the main peaks. 
The relative strength of the two oscillations is not a robust prediction
of the model since it depends on the relative amplitudes of the $\theta$ and 
$r$ oscillations. These can only be predicted by a self-consistent dynamical
simulations.
\label{fig:psd}}
\end{figure}

We note first that under the influence of relativistic effects the shape of
a simple power law spectrum does not change. In the context of our models
this means that a QPO energy spectrum would be the same as the input 
power law spectrum. However, X-ray spectra of X-ray binaries 
are not simple power laws. Moreover, 
the modulation affects only a fraction of emission originating around
the resonance radius, and we can also expect discrete spectral features
(the Fe K$\alpha$ line) to be distorted by these effects. Thus, 
the QPO spectrum can be expected to be different than the time average
spectrum, giving a potential diagnostic of the QPO model. 

\begin{figure*}
 \parbox{\textwidth}{
  \epsfxsize = 0.9\textwidth
  \epsfbox[18 440 620 710]{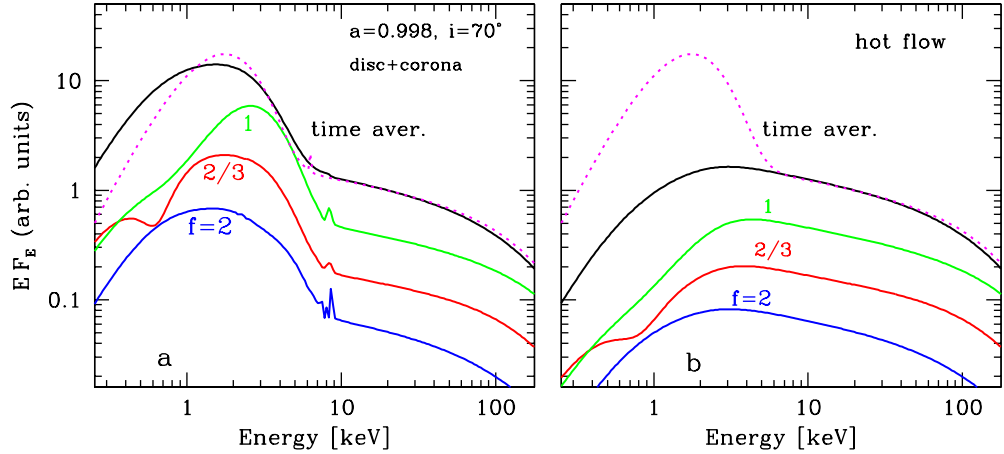}
}
 \caption{The QPO spectra for a maximally rotating black hole case,
for inclination of $70^{\circ}$. The upper thick solid curve (black) is the
time average spectrum of the time dependent component of radiation emitted at 
$\rQPO$ (thus it does not contain the constant disc component in the hot flow
geometry in b). The dotted curve 
(magenta online) is the spectrum from $\rQPO$ but without the relativistic effects.
The three curves labeled $f=2,\ 2/3,\ 1$ (blue, red, green online) 
show the QPO spectra for given $f$ in units of $1/P_\theta$, i.e.\ 
the inverse of oscillation period in the $\theta$-coordinate. 
Left plot (a) shows the spectra for disc with a corona geometry, where the 
relativistic effects affect all spectral components (disc emission, 
Comptonization and the Fe K$\alpha$ line), while plot (b) shows results for
the inner hot flow geometry, where only the Comptonized emission is
modulated.
Normalizations of the spectra are arbitrary, but the order of the QPO curves
corresponds to the strength of the QPO.
\label{fig:ffsp70}}
\end{figure*}

\begin{figure*}
 \parbox{\textwidth}{
  \epsfxsize = 0.9\textwidth
  \epsfbox[18 490 620 710]{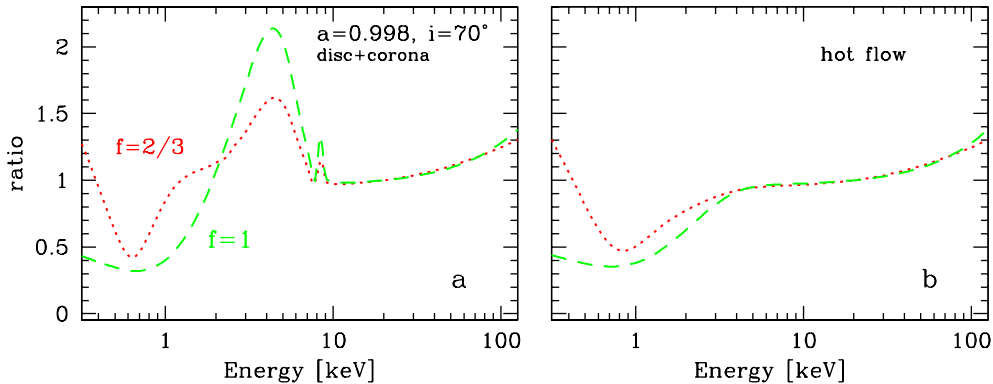}
}
 \caption{Ratios of the QPO spectra for $a=0.998$ and $i=70^\circ$ to
 the time averaged spectra from $\rQPO$ in the disc with a corona geometry (a) and 
the hot flow geometry (b). The QPO spectra are practically identical to
the time averaged spectra above 10 keV. Below that, in the  disc plus corona
geometry the QPO spectra show features related to Fe \Ka line variability, 
while in hot flow geometry QPO spectra show stronger cutoff compared to
the time averaged spectra (see also Fig.~\ref{fig:ffsp70}).
\label{fig:ffrat}}
\end{figure*}

We first discuss the geometry of a disk extending to the marginally stable
orbit, with the hard X-rays produced in a corona extending over a range of 
radii and X-ray reprocessing taking place locally. We assume that both the
disc and the corona undergo epicyclic oscillations and, in consequence, 
the relativistic effects from the deformed precessing ring affect all three 
spectral components. Thus, all three components 
can be expected to appear in the QPO spectrum. Power spectrum
from the light curve at 10 keV, for a maximally rotating black hole with 
$a=0.998$ and a high inclination of $i=70^{\circ}$,  
is plotted in Fig.~\ref{fig:psd}. It shows
the two main QPO peaks, at $f=2/3\ {\rm and} 1\,P_\theta^{-1}$, as well as many other 
(sub)harmonics. Relative height of the two main peak is not a robust 
prediction of the model, since it depends on the (arbitrary) oscillation
amplitudes of the two epicyclic motions.
We note that the other peaks are much weaker than the
two main QPOs, and therefore would be impossible to detect in current data.

\begin{figure*}
 \parbox{\textwidth}{
  \epsfxsize = 0.9\textwidth
  \epsfbox[18 440 620 710]{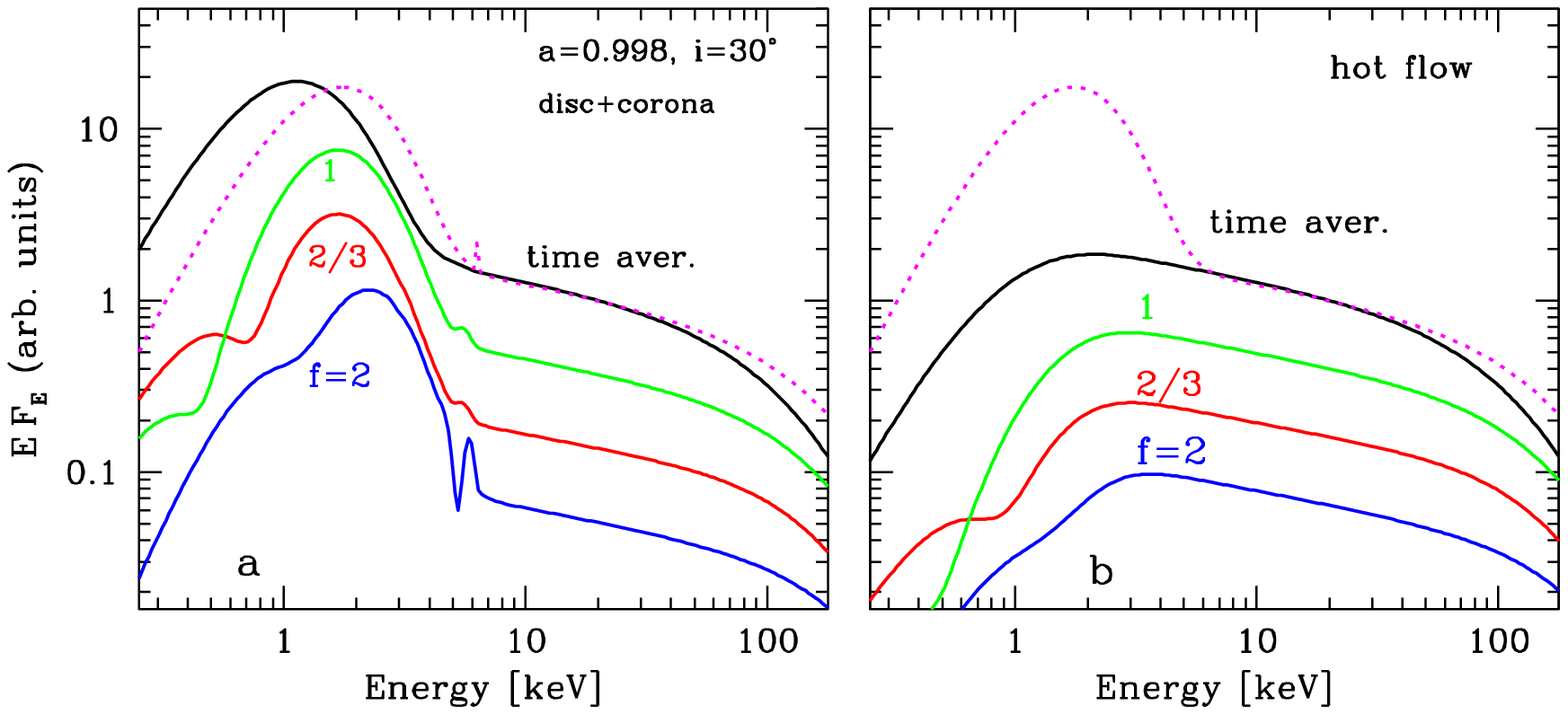}
}
 \caption{The QPO spectra for a maximally rotating black hole case,
for inclination of $30^{\circ}$. The curves are labeled as in 
Fig.~\ref{fig:ffsp70}. 
\label{fig:ffsp30}}
\end{figure*}

\begin{figure*}
 \parbox{\textwidth}{
  \epsfxsize = 0.9\textwidth
  \epsfbox[18 440 620 710]{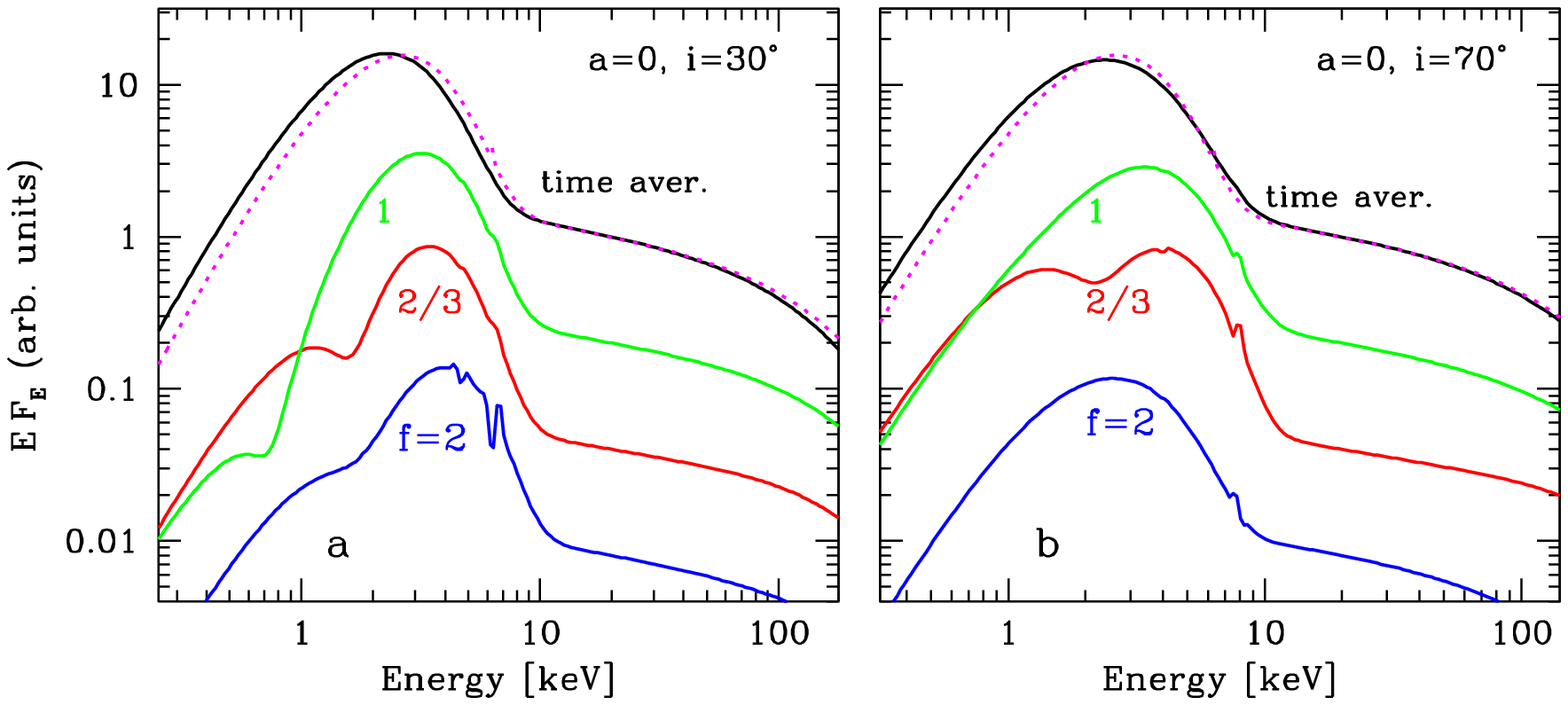}
}
 \caption{The QPO spectra for a non-rotating black hole case 
in the disc with corona geometry, for inclinations of $i=30^{\circ}$ (a) and 
$i=70^\circ$ (b). The curves are labeled as in 
Fig.~\ref{fig:ffsp70}. 
\label{fig:ffspa0}}
\end{figure*}

The QPO spectra for $f=2/3, 1\ {\rm and}\ 2\, P_\theta^{-1}$ 
are plotted in Fig.~\ref{fig:ffsp70}a, together with the
time average spectra for comparison. The QPO spectra do indeed contain
both the disc component and the Fe \Ka line. The disc component in 
the strongest, $f=1$, QPO is rather narrow and significantly blueshifted.
The disc component in the weaker, $f=2/3$, QPO is somewhat broader,
and less strongly blueshifted. In the weak $f=2$ QPO the disk component
is not really distinctly visible and the low energy cutoff is rather
broad. The positions of the Fe \Ka line give us a clue to the behaviour
of the disc component. In all QPO spectra there is a line component around
$E \approx 8.3$ keV, which is the line energy blueshifted by the maximum
blueshift, $\gmax$ for the considered set of parameters: $a$, $\rQPO$ and $i$. 
In the $f=2/3$ and $f=2$ QPOs there are also weak structures around 
$E\approx 2.2$ keV, corresponding to the maximum redshift, $\gmin$. 
Values of these extreme shifts vary as a result of disc precession, and
this causes the observed disc emission to vary. It is the variable
spectrum that appears as the QPO spectrum. The variability is strongest near 
the extreme energy shifts, and therefore discrete spectral features in the 
QPO spectra are mainly located at those extreme positions. Differences
between spectra at different $f$ result from different behaviour
of the extreme energy shifts when folded (observed) at different sub-periods 
($1 P_\theta$ and $P_r = 3/2 P_\theta$) of the total period, $3 P_\theta$.

The QPO spectra in the geometry of inner hot flow are plotted in 
Fig.~\ref{fig:ffsp70}b, for the same $a$ and $i$, for direct comparison.
Here, only the Comptonized component is variable, so neither the disc
nor the Fe line are visible in the QPO spectra. The spectra
are featureless and the only difference between different $f$'s is
in the low energy cutoff, again related to different behaviour 
of energy shifts at different periods.

Fig.~\ref{fig:ffrat} shows ratios of the QPO spectra to the time averaged
spectrum, for $a=1$ and $i=70^\circ$, to demonstrate better the differences
between them. The QPO spectra are different from the mean spectra only
below $\approx 10$ keV. In the disc plus corona geometry they show spectral
features related to the variability of the  Fe \Ka line, as discussed above.
In the hot flow geometry the QPO spectra show stronger cutoffs around 10 keV,
clearly visible in the ratio representation.

The strength of relativistic distortions depends on the inclination angle
of the source, therefore we show in Fig.~\ref{fig:ffsp30} similar
spectra for an inclination of $i=30^\circ$ and $a=0.998$. For the lower $i$
the range of redshift attained is narrower than for a higher $i$, and 
$\gmax<1$, due to strong gravitational redshift.
As a result, the QPO spectra contain a distinct black body component 
(it is less smeared than in the previous case), while the blue component
of the Fe \Ka line is at lower energy than it is in the time averaged spectra. 
The QPO spectra from the hot flow geometry are again featureless, and the
low-energy cutoffs are somewhat sharper than in the high-$i$ case, 
due to narrower range of energy shifts.

Strength of the relativistic modulations depends also on the black hole
spin. There is a dependence on $a$ for a given radius, and, in the context
of our models, there is also a dependence of $\rQPO$ on $a$. We have therefore
computed the QPO spectra for a non-rotating black hole, $a=0$, in the 
disc-corona geometry, for two values of the inclination angle. Results are
shown in Fig.~\ref{fig:ffspa0}. These show strong disc component, as expected
from the geometry, broadened and/or shifted according to the variations of
the extreme energy shifts.

Models computed above demonstrate that the QPO spectrum is generally similar
to the spectrum emitted at the QPO radius. This is reasonable, since 
the relativistic effects simply smear out the emitted spectrum. Therefore,
we also investigated a scenario, where the spectrum emitted at $\rQPO$ is 
rather different than spectra emitted at neighboring radii. This is also
motivated by results of modeling of many high luminosity states spectra,
which reveal additional components. For example, often a very high state
spectrum contains a disc thermal component and {\em two\/} Comptonized 
components, one of which is from relatively cool optically thick plasma.
Studies performed by, e.g., Done \& Kubota (2006) suggest that such a cool
Comptonized  component may originate in inner part of accretion flow.

\begin{figure}
 \parbox{0.5\textwidth}{
  \epsfxsize = 0.45\textwidth
  \epsfbox[18 150 620 710]{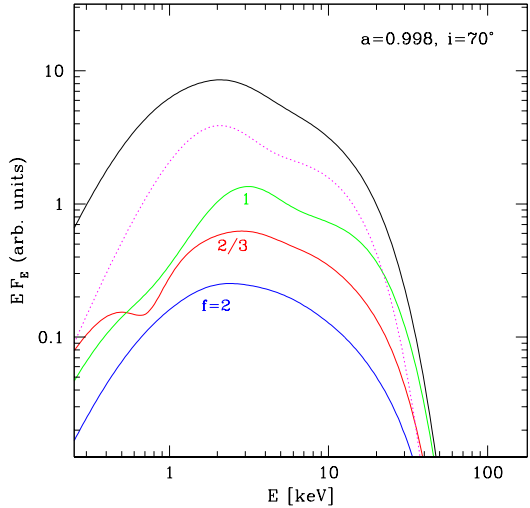}
}
 \caption{QPO spectra from a model, where the energy spectrum emitted
 at $\rQPO$ is a Thomson thick low-$\kT$ Comptonization (dotted curve,
magenta online). The QPO spectra are similar to the spectrum emitted
at $\rQPO$. Curves are labeled as in Fig.~\ref{fig:ffsp70}.
\label{fig:compt}}
\end{figure}

We have therefore assumed that the spectrum produced at $\rQPO$ is
a result of Comptonization of disc photons in a plasma of $\kT \approx 3$ keV and
$\tauT = 7$, while at other radii the parameters are as before.
Results, for $a=0.998$ and $i=70^\circ$, are presented in Fig.~\ref{fig:compt}.
The QPO spectra repeat the pattern observed also in Fig.~\ref{fig:ffsp70}
for the same $a$ and $i$, in the sense that the $f=2$ QPO spectrum is broadest,
and the $f=1$ and $f=2/3$ spectra are blueshifted.

\section{Discussion and conclusions}
\label{sec:discuss}

We have computed QPO energy spectra from the model assuming that the
X-ray modulation is produced by variable relativistic effects from
oscillating accretion flow. The model was suggested 
for the high-frequency QPO observed in a number of black hole X-ray binaries,
where pairs of QPO at frequencies with 3:2 ratio are often observed
(Klu\'{z}niak \& Abramowicz 2001).
In its basic form the model assumes that the intrinsic luminosity
of the X-ray source is constant (or at least there is no periodic component), 
and the entire
observed modulation comes from the relativistic effects. This is currently
the only model of high-$f$ QPO, which actually proposes a well defined
mechanism of modulation of X-rays and thus can make predictions for
spectral observables (see also Giannios \& Spruit 2004 for a model
of low-$f$ QPO).

Observations of hfQPO are rather difficult, but current data
suggest that the thermal disc emission does not participate in the
modulations (see review of observations in Sec.~\ref{sec:data}). 
It is however impossible to determine whether the QPO spectra are 
the same as the harder Comptonization component or they are systematically 
harder or softer than this. 
The former case would mean that the QPO might correspond  to a 
simple modulation of luminosity of the Comptonization source, while the
latter would imply either a dependence of spectral shape on radius, or
spectral variability (e.g., \.{Z}ycki \& Sobolewska 2005).
The lack of modulation of the disc emission is, nevertheless, interesting
since this appears to be a common property between the high-$f$ and
low-$f$ QPO (Sobolewska \& \.{Z}ycki 2006).

Our computations show that the proposed X-ray modulation mechanism
produces QPO spectra which are qualitatively similar to the spectrum
of emission from the QPO radius. Quantitatively, the QPO spectrum are
broadened and shifted under the influence of the Doppler and 
gravitational energy shifts. The amount of smearing depends on the extreme
values of energy shifts, which themselves are functions of black hole spin
(both directly and through the QPO radius, given by the resonance
condition), and the inclination angle. Since the QPO energy spectra are
really variability spectra, they depend also on the way $\gmin$ and 
$\gmax$ vary during the QPO cycle. This causes differences between
spectra of different QPO.

In all our models where the disc thermal component is emitted at the QPO radius,
it is variable and, obviously, it is present in the QPO spectrum. 
If, however, the lack of the disc component is indeed a 
robust feature of the hfQPO spectra, it suggests (within the considered class
of models) that 
the optically thick accretion disc is absent in the region of QPO generation,
and only hot plasma is present. This would support a geometry with an inner
hot torus, perhaps consistent with recent MHD simulations (e.g., Hawley \& 
De Villiers, 2004 and references therein).

Energy dependencies of the two main QPO in the considered model are generally 
very similar, apart from the small differences discussed above. This might
be a test of the model, if it turned out that the observed dependencies
are different. For example, the pair of hfQPO in GRO~J1655-40 
seem to have different energy dependencies: the 450 Hz QPO has
a hard spectrum (rms increasing with $E$), while the 300 Hz seems to have
a soft spectrum. The data are, however, not fully conclusive yet 
(Sec.~\ref{sec:data}).

\section*{Acknowledgments} 
 
This work  was partly supported by grant no.\ 2P03D01225
from the 
Polish Ministry of Science and Higher Education.

{}



\begin{thebibliography}{}
 \bibitem[]{}
 Abramowicz M. A., Karas V., Klu\'{z}niak W., Lee W. H., Rebusco P.,
  2003, PASJ, 55, 467
 \bibitem[]{} Bardeen J. M., Press, W. H., Teukolsky, S. A., 1972,
  ApJ, 178, 347
 \bibitem[]{}
  Belloni T., Soleri P., Casella P., Mendez M., Migliari, S., 2006, 
            MNRAS, 369, 305

 \bibitem[]{}
  Blaes O. M., Arras P., Fragile P. C., 2006, MNRAS, 369, 1235
 \bibitem[]{}
  Bursa M., Abramowicz M. A., Karas V., Klu\'{z}niak W., 2004, ApJ, 617, L45
 \bibitem[]{}
   Cui W., 1999, ApJ, 524, L59
 \bibitem[]{}
   Cui W., Shrader C. R., Haswell C. A., Hynes R. I., 2000, ApJ, 535, L123
 \bibitem[]{}
   Done C., Kubota A., 2006, MNRAS, 371, 1216
 \bibitem[]{}
   Giannios, D.; Spruit, H. C, 2004, A\&A, 427, 251
 \bibitem[]{} 
   Gierli\'{n}ski M., Done C., 2003, MNRAS, 342, 1083
 \bibitem[]{} 
 Hawley J. F., De Villiers J.-P., 2004, PThPS, 155, 132   
 \bibitem[]{} 
  Homan J., Wijnands R., van der Klis M., Belloni T., van Paradijs J.,
   Klein-Wolt M., Fender R.,; Mendez M., 2001, ApJS, 132, 377
\bibitem[]{} 
 Homan J., Klein-Wolt M., Rossi S., Miller J. M., Wijnands R., Belloni T.,
   van der Klis M., Lewin W. H. G., 2003, ApJ, 586, 1262
 \bibitem[]{} 
  Homan J., Miller J. M., Wijnands R., van der Klis M., Belloni T.,
   Steeghs D., Lewin W. H. G., 2005, ApJ, 623, 383
\bibitem[]{}
  Kalemci E., Tomsick J. A., Rothschild R. E., Pottschmidt K., Kaaret P.,
   2001, ApJ, 563, 239
\bibitem[]{}
    Klein-Wolt M., Homan J., van der Klis M., 2004, NuPhS, 132, 381
\bibitem[]{}
   Klu\'{z}niak W., Abramowicz M. A., 2001, AcPPB, 32, 3605
\bibitem[]{}
   Krolik J. H., 1999, Active Galactic Nuclei, Princeton University Press,
    Princeton
\bibitem[]{}
  Lachowicz P., Czerny B., Abramowicz M., 2006, MNRAS, submitted, astro-ph/0607594
 \bibitem[]{}
  Miller J. M. et al., 2001, ApJ, 563, 928 
\bibitem[]{}
   Morgan E. H., Remillard R. A., Greiner J., 1997, ApJ, 482, 993
\bibitem[]{}
   Nied\'{z}wiecki A., 2005, MNRAS, 356, 913

 \bibitem[]{}
   Remillard R. A., McClintock J. E., 2006, ARA\&A, 44, 49
 \bibitem[]{}
   Remillard R. A., McClintock J. E., Sobczak G. J., Bailyn C. D., 
  Orosz J. A., Morgan E. H., Levine A. M., 1999a, ApJ, 517, L127
 \bibitem[]{}
   Remillard R. A., Morgan E. H., , McClintock J. E., Bailyn C. D., 
  Orosz J. A., 1999b, ApJ, 522, 397
  \bibitem[]{}
   Remillard R. A., Muno M. P., McClintock J. E., Orosz J. A., 
     2002, ApJ, 580, 1030
 \bibitem[]{}
   Remillard R. A., McClintock J. E., Orosz J. A., Levine A. M., 
    2006, ApJ, 637, 1002
 \bibitem[]{}
   Sobolewska M. A.,  \.{Z}ycki P. T., 2006, MNRAS, 370, 405
 \bibitem[]{}
   Strohmayer T. E., 2001a, ApJ, 552, L49
 \bibitem[]{}
   Strohmayer T. E., 2001b, ApJ, 554, L169
\bibitem[]{}
   T\"{o}r\"{o}k G., Abramowicz M. A., Klu\'{z}niak W., Stuchl\`{i}k Z.,
 2004, A\&A, 436, 1

 \bibitem[]{}
   Zdziarski A. A., Johnson W. N., Magdziarz P., 1996, MNRAS, 283, 193
 \bibitem[]{}
    \.{Z}ycki P. T., 2003, MNRAS, 340, 639
 \bibitem[]{}
    \.{Z}ycki P. T., Nied\'{z}wiecki A., 2005, MNRAS, 359, 308
 \bibitem[]{}
    \.{Z}ycki P. T., Sobolewska M. A., 2005, MNRAS, 364, 891

\label{lastpage}

\end{thebibliography}
\end{document}